\documentclass[twocolumn,prl,amsmath,amssymb,superscriptaddress]{revtex4} 
\usepackage{graphicx,macros,color,pdfpages}

\begin{document}
\title{Emergence of information transmission in a prebiotic RNA reactor}
\author{Benedikt Obermayer}\altaffiliation{Present address: Department of Physics, Harvard University, Cambridge MA 02138, USA.}\affiliation{Arnold-Sommerfeld-Center f\"ur Theoretische Physik and Center for NanoScience, and}
\author{Hubert Krammer} \affiliation{Systems Biophysics, Physics Department, Center for Nanoscience, Ludwig-Maximilians-Universit\"at M\"unchen, 
Germany}
\author{Dieter Braun} \affiliation{Systems Biophysics, Physics Department, Center for Nanoscience, Ludwig-Maximilians-Universit\"at M\"unchen, 
Germany}
\author{Ulrich Gerland}\email{gerland@lmu.de}\affiliation{Arnold-Sommerfeld-Center f\"ur Theoretische Physik and Center for NanoScience, and} 
\begin{abstract}
A poorly understood step in the transition from a chemical to a biological world is the emergence of self-replicating molecular systems. We study how a precursor for such a replicator might arise in a hydrothermal RNA reactor, which accumulates longer sequences from unbiased monomer influx and random ligation. In the reactor, intra- and inter-molecular basepairing locally protects from random cleavage. By analyzing stochastic simulations, we find temporal sequence correlations that constitute a signature of information transmission, weaker but of the same form as in a true replicator.
\end{abstract}

\maketitle

The RNA world theory~\cite{GilbertNature:86} posits that the first information carrying and catalytically active molecules at the origin of life were RNA-like polynucleotides \cite{OrgelCRBM:04}. This idea is empirically supported by the discovery of ribozymes, which perform many different reactions~\cite{DoudnaNature:02}, among them the basic template-directed ligation and polymerization steps~\cite{PaulPNAS:02, JohnstonScience:01} necessary for replicating RNA. However, a concrete scenario how a self-replicating RNA system could have arisen \emph{spontaneously} from a pool of random polynucleotides is still lacking. Physical effects may have facilitated this step, as is believed to be the case in other transitions of prebiotic evolution \cite{ChenScience:04}. 

\begin{figure}[b]
\includegraphics[width=.45\textwidth]{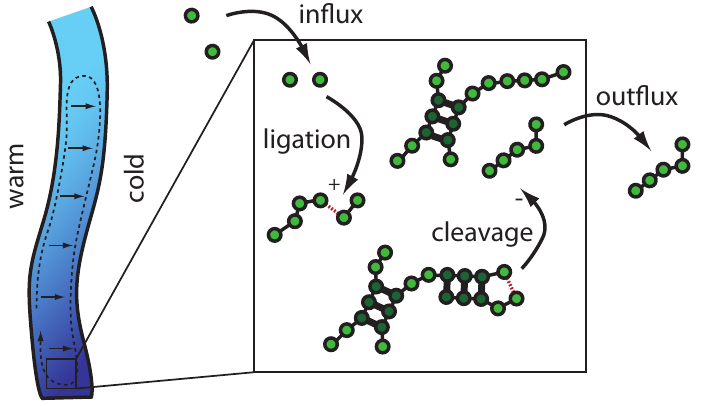}
\caption{\label{fig:schema}(Color online) Illustration of the RNA reactor. Left: Combined action of convection and thermophoresis in narrow pores subject to a temperature gradient results in strong accumulation of nucleotides, as indicated by the darker shading. Right: The region of high concentration defines an open reaction volume where nucleotides enter and bonds are formed through ligation reactions. Equilibrium base pair formation protects bonds next to paired nucleotides (dark) from cleavage. Length-dependent outflux accounts for the preferential accumulation of long molecules.}
\end{figure}

From the perspective of information, an RNA replicator transmits sequence information from molecule to molecule, such that the information survives even when the original carrier molecules are degraded, for instance due to hydrolytic cleavage~\cite{UsherPNAS:76}. Rephrased in these terms, the problem of spontaneous emergence of an RNA replicator \cite{FernandoJME:07, NowakPNAS:08} becomes a question of a path from a short term to a lasting sequence memory. Either this transition occurred as a single unlikely step or as a more gradual, multi-step transition. Here, we explore a scenario of the latter type, based only on simple physico-chemical processes, see Fig.~\ref{fig:schema}: (i) random ligation of RNA molecules, e.g. in a hydrothermal ``RNA reactor'', where polynucleotides are accumulated by thermophoresis~\cite{BaaskePNAS:07}, (ii) folding and hybridization of RNA strands, and (iii) preferential cleavage of single- rather than double-stranded RNA segments \cite{UsherPNAS:76}. Using extensive computer simulations and theoretical analysis, we study the behavior that emerges when these processes are combined. 

Clearly, the preferential cleavage at unpaired bases effectively creates a selection pressure for base pairing in the reactor. We find that this effect increases the complexity of RNA structures in the sequence pool, which may favor the emergence of ribozymes. The underlying sequence bias also extends the expected lifetime of sequence motifs in the finite pool. Interestingly, we find that correlations between motifs persist even longer than expected. This memory effect is associated with information transmission via hybridization. Intriguingly, these correlations have the same statistical signature as templated self-replication, only weaker. In this sense, the RNA reactor could constitute a stepping-stone from which a true RNA replicator could emerge, e.g., assisted by a primitive ribozyme catalyzing template-directed synthesis. 

\paragraph{RNA reactor.---}
As illustrated in Fig.~\ref{fig:schema}, we envisage an open reaction volume $V$ under non-equilibrium conditions as, e.g., inside a hydrothermal pore system where polynucleotides are strongly accumulated by a combination of convective flow and thermophoresis~\cite{BaaskePNAS:07}. At any point in time, the reaction volume contains various sequences $S_L$ of length $L$. The full time evolution of this pool is a stochastic process with the reactions 
\begin{subequations}\label{eq:reactions}
\begin{align}
\emptyset &\stackrel{J}{\longrightarrow} S_1 &
S_L &\stackrel{d_L}{\longrightarrow} \emptyset \\
S_L+S_K &\stackrel{\alpha}{\longrightarrow} S_{L+K} &
S_L &\stackrel{\beta_{L,K}}{\longrightarrow} S_K + S_{L-K}\;.
\end{align}
\end{subequations}
We assume a constant and unbiased influx of monomers (ACGU) at rate $J$. The effective outflux rate $d_L=d_0 \e^{-(L/L_\text{c})^{1/2}}$ accounts for the strong accumulation of nucleotides in a pore system, with a characteristic length dependence determined by the length scale $L_\text{c}$, which comprises parameters such as Soret coefficient, temperature gradient, and geometry \cite{DuhrPNAS:06}.  Ligation of monomers or oligomers occurs at fixed rate $\alpha$~\cite{Note1}. Finally, the most essential ingredient is a backbone cleavage process with a rate that depends on the base-pairing probability of the neighboring bases, such that double-stranded RNA is more stable than single-stranded RNA. Specifically, we calculate the cleavage rate $\beta_{L,K}=\beta_0(1-p_{L,K})$ at backbone bond $K$ using the average base-pairing probability $p_{L,K}$ of the two neighboring bases. We allow both intra-molecular base pairs within single sequences and inter-molecular base pairs within duplexes of any two molecules. RNA folding is performed by means of the Vienna package~\cite{HofackerMC:94,Note2}, where the partition function of the entire ensemble is calculated assuming chemical equilibrium~\cite{BernhartAMB:06}, warranted by the fast hybridization kinetics~\cite{FernandoJME:07}.

We use the standard Gillespie algorithm to simulate the stochastic dynamics (\ref{eq:reactions}) of the sequence pool. The cleavage rate $\beta_{L,K}$, which is recalculated from the folding output for all molecules whenever necessary, effectively introduces a selection for base-pair formation. Since RNA folding depends on the temperature $T$ and duplex formation is also concentration-dependent, we can vary the selection pressure via $p_{L,K}(T,V)$. We consider the reactions (\ref{eq:reactions}) under different possible conditions, with two different temperatures (a cold system at $10^\circ$C and a hot environment at $60^\circ$C) and concentrations (in the pM and mM range, respectively). To study the differences to a random pool, we also consider a ``neutral'' scenario without folding ($p_{L,K}=0$). These scenarios are chosen mainly to highlight the effects of base pairing and not to suggest specific environmental conditions at the origin of life.

\begin{figure}
\includegraphics[width=.48\textwidth]{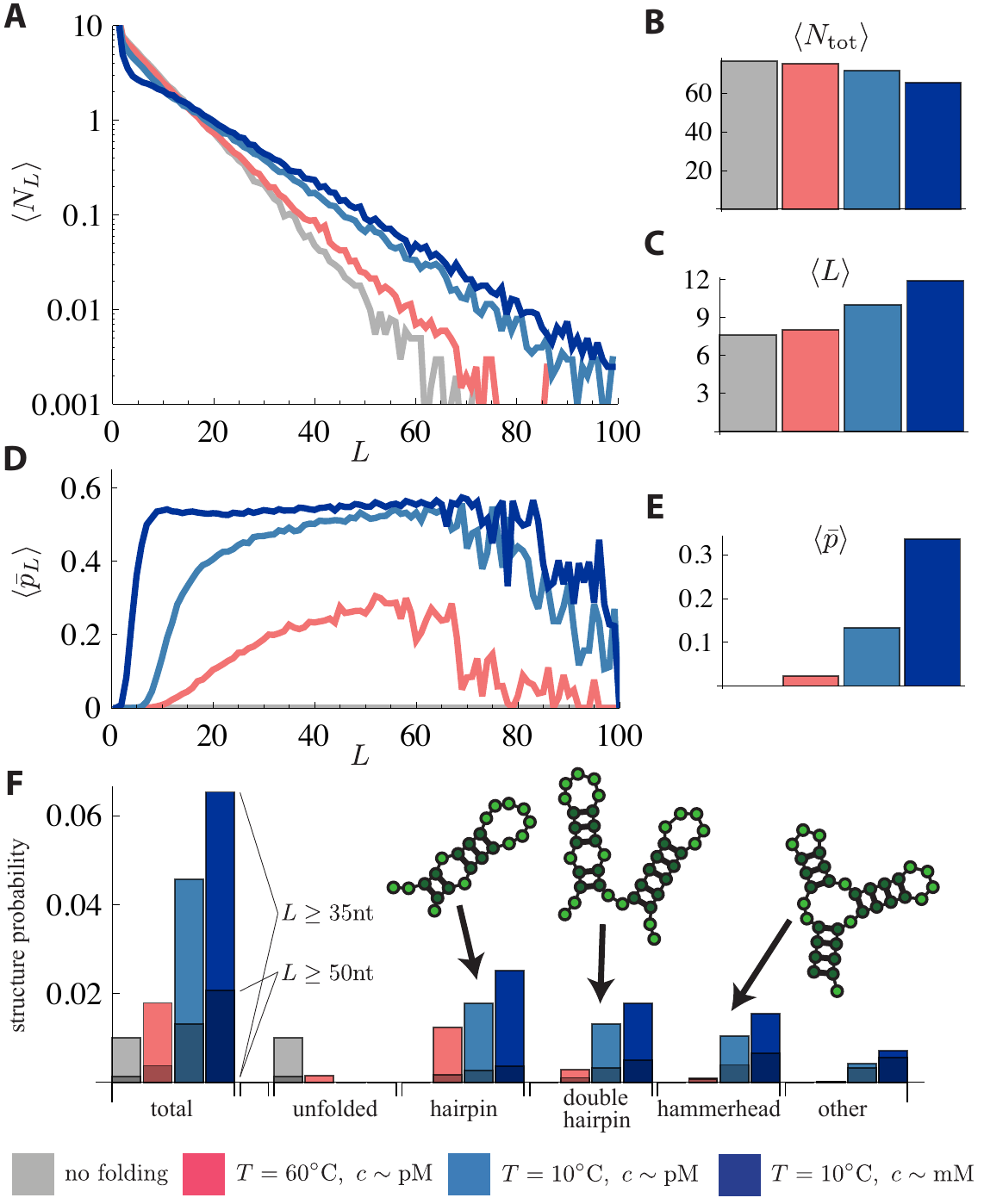}
\caption{\label{fig:stats}(Color online) Steady-state properties of the sequence pool: (a) length distribution $\ave{N_L}$, (b) total number $\ave{N_\text{tot}}$ of molecules, and (c) their mean length $\ave{L}$. (d) base pairing probability $\ave{\bar p_L}$ averaged over sequences of length $L$, with mean $\ave{\bar p}$ shown in (e). (f) structural repertoire of long sequences: steady-state probabilities for sequences longer than $L^*=35$ (shaded parts: $L^*=50$), which fold into a structure of similar shape as the indicated schematic drawings. Selection strength increases from light to dark color as indicated in the legend. All observables are averaged over time and 10 independent replicas. Remaining parameter values were $J=1$, $\alpha=.001$, $\beta_0=.01$, $d_0=.005$, $L_\text{c}=10$.}
\end{figure}

\paragraph{Stationary length and shape distribution.---}
Disregarding sequence-dependent selection, the ligation-cleavage dynamics of the RNA reactor resembles the kinetics of cluster aggregation and fragmentation. Hence, the stationary sequence length distribution shown in Fig.~\ref{fig:stats}(a) corresponds to a cluster size distribution, and its moments can be obtained using established methods~\cite{LiJ:05,Note2}. In the limit of large influx $J$, the average total molecule number $\ave{N_\text{tot}}$ and their mean length $\ave{L}$ are given by: 
\begin{equation}\label{eq:moments}
\ave{N_\text{tot}} = \sqrt{\frac{J (d_0+\beta_0)}{\alpha d_0}}, \quad \ave{L}=\sqrt{\frac{J\alpha}{d_0(\beta_0+d_0)}}\;,
\end{equation}
where we have neglected the length dependence of the outflux ($L_\text{c}\to\infty$; a finite value for $L_\text{c}$ shifts both $\ave{N_\text{tot}}$ and $\ave{L}$ to larger values without strongly affecting the shape of the distribution). These analytical results readily explain why with stronger selection the total number of molecules decreases, but their mean length goes up (see Fig.~\ref{fig:stats}(b) and (c)): the cleavage rate $\beta_0$ is reduced as the mean base pairing probability $\ave{\bar p_L}$ is increased especially for longer sequences (cf.\ Fig.~\ref{fig:stats}(d)), and the distribution thus gains more weight in the tail of long sequences. 

In order to characterize the structural repertoire of this RNA pool, we focused on the tail of the length distribution and analyzed the secondary structures of long sequences with $L > L^*$. We performed the analysis for $L^*=35$ as well as $L^*=50$ (the length of the minimal hairpin ribozyme~\cite{HampelBiochemistry:89}). Fig.~\ref{fig:stats}(f) shows the probability to observe structures within basic ``shape'' classes~\cite{GiegerichNAR:04}, such as hairpins or hammerheads~\cite{Note3}. We observe a significant enrichment of complex structures under selection compared to the neutral case defined above.

\begin{figure}
\includegraphics[width=.48\textwidth]{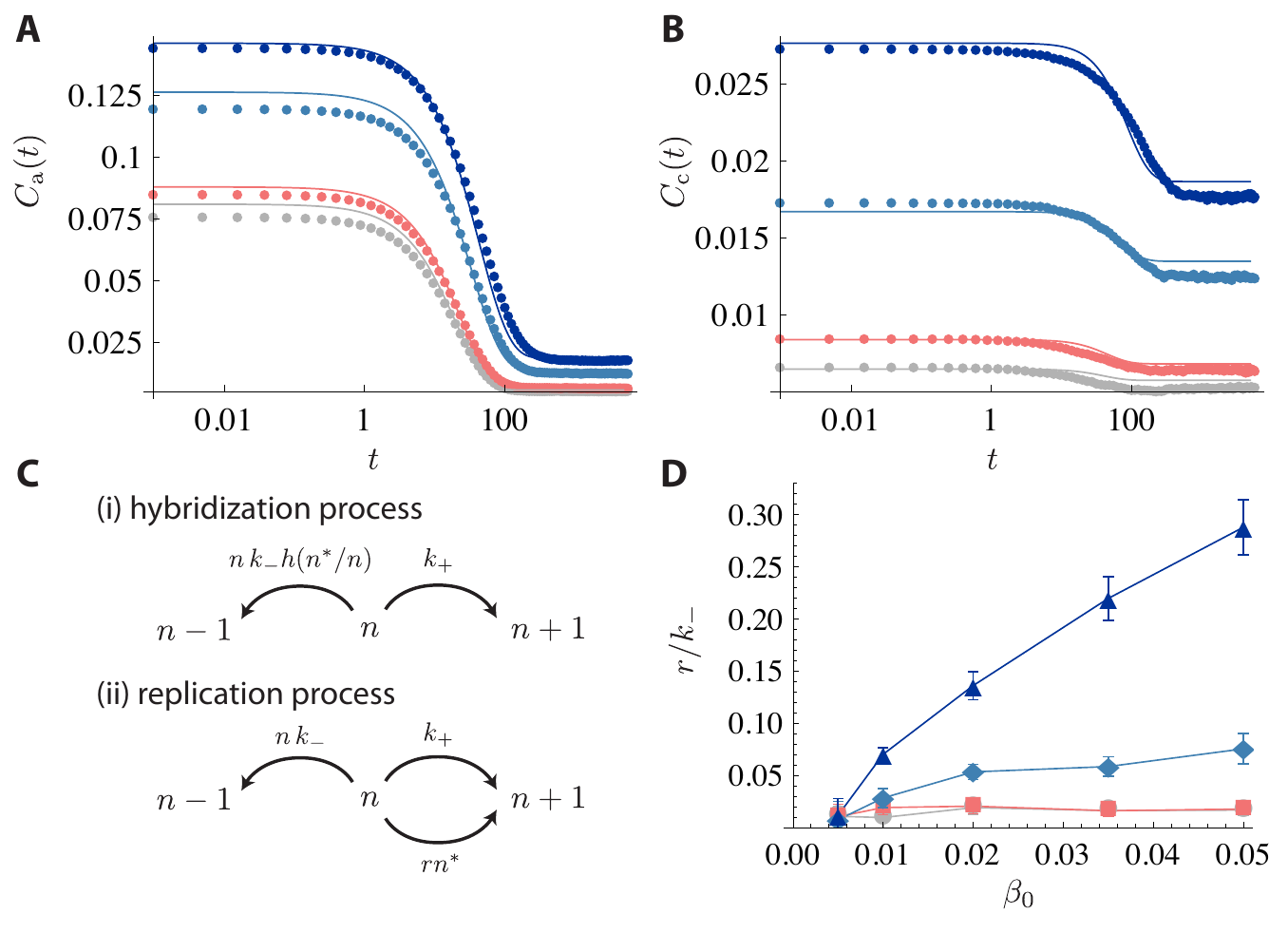}
\caption{\label{fig:replication}(Color online) Information transmission among sequence motifs. (a) and (b) auto- and cross-correlation functions $C_\text{a/c}(t)$ from simulation data for $\ell=6$ (dots) together with analytical expressions from Eq.~\eqref{eq:correlation-functions} (solid lines). The rates $k_-$ and $k_+$ have been computed from Eqs.~\eqref{eq:kp} and \eqref{eq:km}, with $r$ as only fit parameter. (c) schemata for different birth-death processes: (i) motifs are created with constant rate $k_+$ and destroyed with linear rate $k_- h(n^*/n)$, which is reduced by hybridization to their complements; (ii) motifs are destroyed with fixed rate $k_-$, but are copied from their complements with rate $r$. To leading order in $r/k_-$, both processes give rise to identical correlation functions $C_\text{a/c}(t)$, where a nonconstant $C_\text{c}(t)$ signifies information transmission between a motif and its complement. (d) dependence of the replication efficiency $r/k_-$ on the cleavage rate $\beta_0$ (error bars indicate $95\,\%$ confidence intervals). Color code as in Fig.~\ref{fig:stats}.}
\end{figure}

\paragraph{Information transmission via hybridization.---}  
Base pairing and the ensuing correlations between sequences occur mostly within relatively short sequence regions. Therefore, we focus on the dynamics of shorter subsequences or ``sequence motifs'' of length $\ell$, which are informational entities not tied to a specific molecule. From our simulations, we extract time trajectories for the copy numbers $n_i(t)$ of all $4^\ell$ different motifs. Even for fairly small $\ell > 3$, the sequence space of motifs is not fully covered in the finite ensemble, i.e., an average motif copy number is typically $\ave{n_i(t)}\ll 1$. Hence, signatures of information transmission should appear as an unexpected increase in the lifetime of these motifs. Suitably averaged observables are provided by the auto- and cross-correlation functions, $C_\text{a}(t) = 4^{-\ell}\sum_i \ave{n_i(t) n_i(0)}$ and $C_\text{c}(t) = 4^{-\ell}\sum_i\ave{n_i(t) n^*_i(0)}$, respectively, where $n_i^*$ is the copy number of a motif's (reverse) complement~\cite{Note3}. Fig.~\ref{fig:replication}(a) and (b) show data for these correlation functions for $\ell=6$ and the parameter set used in Fig.~\ref{fig:stats}.

The observed motif correlations can be understood in the framework of a simple stochastic process. Motifs are created when sequence ends are ligated together and destroyed by cleavage~\cite{Note4}. Using a mean-field-type approach, we pick an arbitrary probe motif with copy number $n(t)$. Its dynamics is described by a birth-death process, where $n(t)$ is increased with constant rate $k_+$ and decreased with linear rate $k_-$, see schema (i) in Fig.~\ref{fig:replication}(c). The birth rate $k_+$ can be computed from the steady-state length distribution $\ave{N_L}$ by counting how many ends of long enough molecules are available for ligation. Assuming an annealed random ensemble, we obtain
\begin{equation}\label{eq:kp}
k_+ = \frac{\alpha}{4^\ell}\sum_{k=1}^{\ell-1} \sum_{L\geq k}\ave{N_L}\!\!\sum_{L'\geq \ell-k}\!\!\ave{N_{L'}}.
\end{equation}
The death rate $k_-$ comprises the effects of cleavage and hybridization. A motif is cleaved with rate $\beta_0$ at any of its $\ell-1$ bonds, but this rate is reduced by the effective base pairing probability of its parent sequence, which in turn depends on the selection strength. On average, this reduction follows from averaging over the length and base-pairing probability distributions $\ave{N_L}$ and $\ave{\bar p_L}$ of parent sequences, respectively. This gives the result
\begin{equation}\label{eq:km}
k_- = \beta_0 (\ell-1)\left[1-\frac{\sum_{L\geq \ell}(L-\ell+1)\ave{\bar p_L}\ave{N_L}}{\sum_{L\geq \ell} (L-\ell+1) \ave{N_L}}\right].
\end{equation}
However, a birth-death process based on these two effective rates alone necessarily fails to describe cross-correlations between a motif and its complement~\cite{Note5}. The reduction in the cleavage rate of a particular motif due to hybridization is \emph{conditional} on the presence of its complementary partner. Hence, we modulate the average death rate $k_-$ with an additional factor $h(x) \leq 1$, which accounts for the probability of hybridization and depends on the number $x=n^*/n$ of available complements per motif. Since the average hybridization probability is small under the conditions considered here, it will be proportional to $x$. This leads us to a linear ansatz $h(x)\approx 1- (r/k_-) x$, where the significance of the coefficient $r$ will shortly become apparent. We find that in the ``hybridization process'' of Fig.~\ref{fig:replication}(c), the expected copy number $\ave{n}$ of a motif obeys 
\begin{equation}\label{eq:nmean}
\pd_t\ave{n}=k_+-k_- \ave{n h(n^*/n)} \approx k_+ - k_- \ave{n} + r \ave{n^*} \;.
\end{equation}
A symmetric equation holds for $\ave{n^*}$. Strikingly, this result is \emph{identical} to the corresponding rate equations for a ``replication process''~\cite{Note2}, where motifs are born with rate $k_+$, destroyed with fixed rate $k_-$, and \emph{copied} from their complements with rate $r$, as in schema (ii) of Fig.~\ref{fig:replication}(c). This observation suggests that we may interpret the coefficient $r$ as an \emph{apparent} replication rate for motifs in the RNA reactor. 

To validate this interpretation, and to measure the apparent replication rate in our simulations, we calculate the correlation functions of the hybridization process using the same approximation for $h(x)$~\cite{Note2}, yielding
\begin{equation}\label{eq:correlation-functions}
C_\text{a/c}(t) = \frac{k_+^2}{(k_--r)^2} + \frac{k_+\e^{-(k_--r) t}}{2(k_--r)} \pm \frac{k_+ \e^{-(k_-+r) t}}{2(k_-+r)}.
\end{equation}
In Fig.~\ref{fig:replication}(a) and (b), we used these expressions with the rates $k_+$ and $k_-$ calculated from Eqs.~\eqref{eq:kp} and \eqref{eq:km}, and with $r$ as only free parameter fitted simultaneously to both datasets. The equivalence between the hybridization and the replication processes is also exhibited by their correlation functions to leading order in $r/k_-$~\cite{Note2}. Hence, the good agreement with the simulation data indicates that the observed motif correlations are virtually indistinguishable from those expected for inefficient template-directed replication. The replication efficiency $r/k_-$ determined by the fits is plotted in Fig.~\ref{fig:replication}(d) as function of the bare cleavage rate $\beta_0$ for the different conditions. Remarkably, it reaches levels close to $30\,\%$ in the cold and highly concentrated environment, where base pairing via duplex formation is favorable. Note that a true (exponential) replicator would require that motifs are copied faster than they are degraded ($r > k_-$), while our system with $r < k_-$ is an inefficient realization.

These findings show that protection against cleavage due to folding and hybridization leads to an extended sequence memory in the RNA reactor. One global contribution to this 
longer motif lifetime is due to the ``protection factor'' in square brackets in Eq.~\eqref{eq:km}, which renormalizes the bare cleavage rate to account for the average probability that a motif is paired. Another contribution stems from the correlation time in Eq.~\eqref{eq:correlation-functions}, which is increased as the apparent replication rate is subtracted from the renormalized cleavage rate, such that $C_\text{a/c}(t)$ decays on time scales of order $(k_--r)^{-1}$. This specific increase occurs only when a motif and its complement mutually protect each other, and it therefore demonstrates the emergence of information transmission.

\paragraph{Conclusions.---} 
We have analyzed stochastic simulations of a minimal prebiotic RNA reactor, where formation of double strands protects sequence parts from degradation. On the one hand, this selection for structure biases the resulting pool towards longer and more structured sequences, favoring the emergence of ribozymes. On the other hand, it leads to a weak apparent replication process based on ``information transmission by hybridization'', conceptually similar to ``sequencing-by-hybridization'' techniques~\cite{DrmanacABEB:02}. Together, the structural complexity and the information transmission featured in the RNA reactor suggests this type of system as plausible intermediate for the emergence of a true replicator with $r > k_-$. For instance, some of the relatively frequent simple structures observed in our simulation are similar to known ligase ribozymes~\cite{DoudnaNature:02}. This functionality in turn would facilitate the creation of more complex molecules from essential modular subunits~\cite{BrionesRNA:09}. Once ribozymes emerge, a self-replicating system could be established by template-directed ligation of suitably complementary oligomers~\cite{PaulPNAS:02}. So far, it remained unclear how such auto-catalytic RNA systems would be supplied with appropriate oligomer substrates. However, the strong cross-correlations observed in the RNA reactor demonstrate a significantly enhanced chance of finding sequences complementary to those present in the pool, including the sequence to be replicated. Thus, the RNA reactor acts as an adaptive filter to preferentially keep potentially useful substrate sequences. This adaptive selectivity would allow for the ``heritable'' propagation of small variations and thus endow the replicator with basic evolutionary potential.

\acknowledgments
This work was supported by the Nanosystems Initiative Munich (NIM), by a DAAD grant to BO, and by a DFG grant to UG.



\section*{Supplementary material}

# Emergence of information transmission in a prebiotic RNA reactor – Supplementary Information

Benedikt Obermayer,[1,*] Hubert Krammer,[2] Dieter Braun,[2] and Ulrich Gerland[1,†]

[1]*Arnold-Sommerfeld-Center für Theoretische Physik and Center for NanoScience,*
*Ludwig-Maximilians-Universität München,*
*Theresienstr. 37, 80333 München, Germany*

[2]*Systems Biophysics, Physics Department,*
*Center for Nanoscience, Ludwig-Maximilians-Universität München,*
*Amalienstr. 54, 80799 München, Germany*

---

[*] Present address: Department of Physics, Harvard University, Cambridge MA 02138, USA.
[†] gerland@lmu.de

# I. SUPPLEMENTARY FIGURES

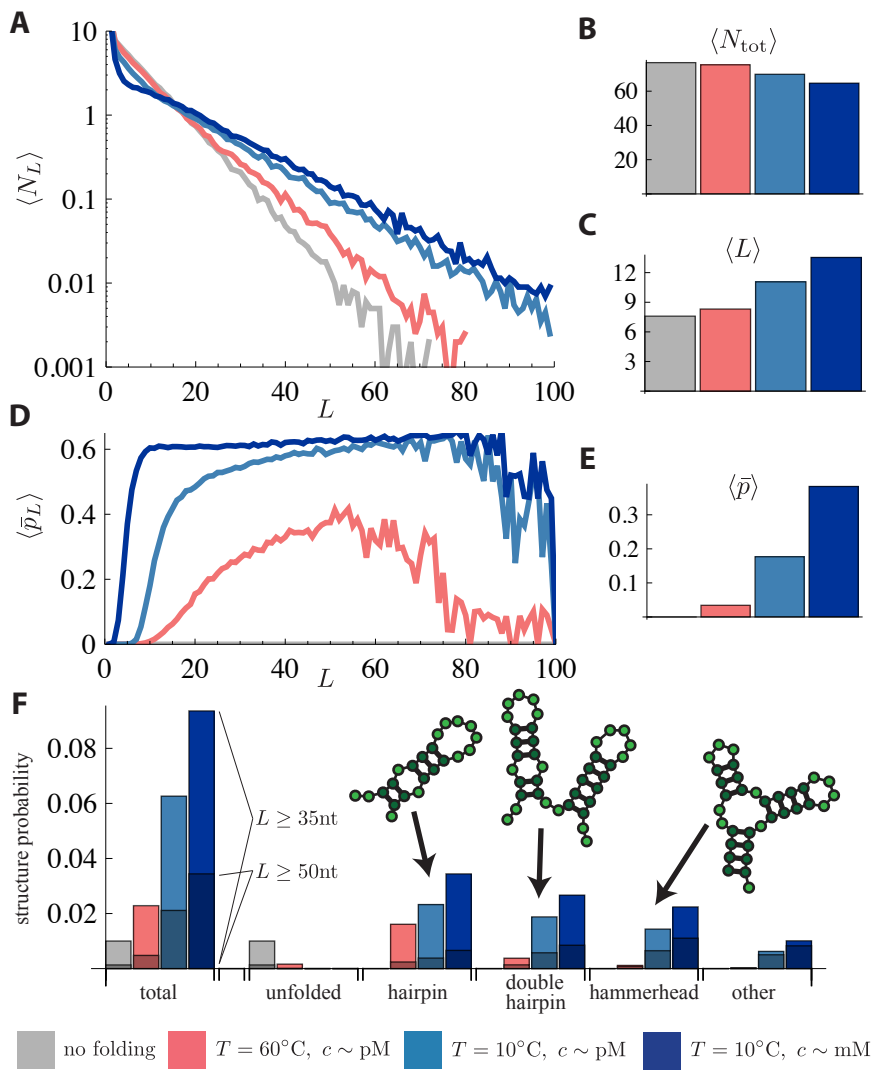

FIG. S1. GU pairs. Properties of the steady-state ensemble as in Fig. (1) of the main text, but in a simulation including GU wobble pairs. (a) length distribution $\langle N_L \rangle$, (b) total number $\langle N_{\text{tot}} \rangle$ of molecules, and (c) their mean length $\langle L \rangle$. (d) base pairing probability $\langle \bar{p}_L \rangle$ averaged over sequences of length $L$, with mean $\langle \bar{p} \rangle$ shown in (e). (f) Structural repertoire of long sequences. While the differences to the results without GU pairs shown in Fig. 2 in the main text are comparably small, we observe that this additional pairing mode provides additional stability especially for longer RNA and thus further increases the chances of finding structured molecules in random pools.

2

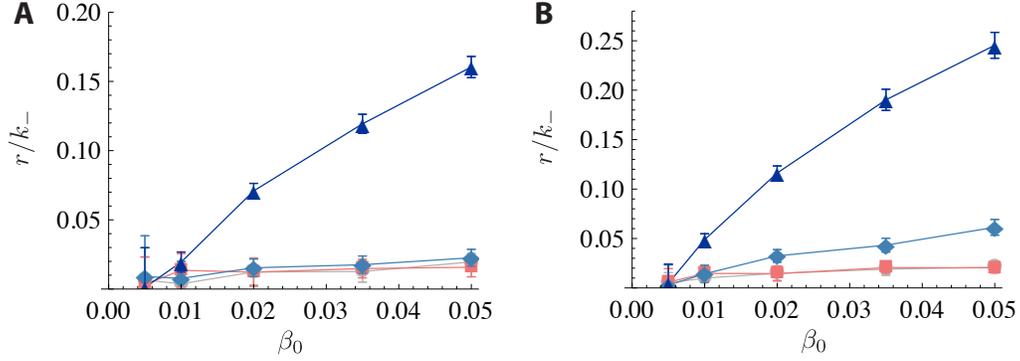

FIG. S2. Shorter motifs. Dependence of the replication efficiency $r/k_-$ on the bare cleavage rate $\beta_0$ as in Fig. (3c) in the main text, but for shorter motifs of length $\ell = 4$ (a) and $\ell = 5$ (b).

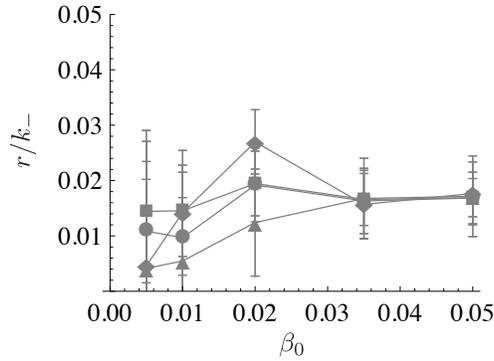

FIG. S3. Analysis of the influence of self-complementary sequences. Self-complementary sequences in the pool give rise to different motif statistics. To test this effect, we ran control simulations without RNA folding but fixed sequence-independent base pairing probabilities $\langle \bar{p}_L \rangle$ chosen from the distributions measured in the full simulation (cf. Fig. 1(d) in the main text). This leads to almost identical length statistics in the sequence ensemble, but motif correlations due to hybridization are absent. Shown is the dependence of the apparent replication efficiency $r/k_-$ on the bare cleavage rate $\beta_0$ as in Fig. (3c) in the main text. Self-complementarity gives rise to subdominant cross-correlations resulting in small non-zero values for $r$ largely independent of the "selection strength" (note the different scale on the ordinate).

## II. IMPLEMENTATION DETAILS: CALCULATION OF BASE PAIRING PROBABILITIES

Our code is based on the Gillespie algorithm [1] for the stochastic simulation of chemical reactions. At each time step, we compute the propensities for each of the four possible reactions involving sequences $S_{L,i}$ of length $L_i$ that are present in $N_{L,i}$ copies:

1. influx of a monomer with propensity $J$ (monomers are chosen randomly among the four nucleotides A,C,G, and U);

2. outflux of one of $N_{L,i}$ copies of $S_{L,i}$ with propensity $d_0 N_{L,i} e^{-\sqrt{L_i/L_c}}$;

3. ligation of two sequences $S_{L,i}$ and $S_{L,j}$ to a combined sequence of length $L_i + L_j \leq L_{\max}$ with propensity $\alpha N_{L,i}(N_{L,j} - \delta_{ij})$.

4. cleavage of one of $N_{L,i}$ copies of a sequence $S_{L,i}$ at position $K$ with propensity $N_{L,i}\beta_{L_i,K}$ where $\beta_{L_i,K} = \beta_0(1 - p_{L_i,K}(T,V))$.

One event is randomly chosen according to its relative propensity, and time is updated by a time interval drawn from an exponential distribution with a mean equal to the inverse of the sum of the propensities.

The first two steps are straightforwardly implemented, but some explanations on the latter two are in order. Firstly, we neglect a possible length dependence of the ligation reaction, which is poorly understood on a microscopic level, but probably rather weak [2]. Also, we scale out its volume dependence to facilitate comparison of different scenarios, which operate at different concentrations. Finally, we restrict ligation to sequences with combined length smaller than $L_{\max} = 100$ to limit computationally expensive RNA folding. Secondly, the cleavage reaction involves the sequence-specific, temperature- and concentration-dependent probability $p_{L_i,K}(T,V)$ that the nucleotides next to bond $K$ are paired. The calculation is done by means of the Vienna package for RNA secondary structure folding [3]. We allow both intra- and intermolecular base pairs in complexes involving at most two sequences. To simplify the following argument, we omit the length index on the sequences $S_i$. For each sequence $S_i$, we calculate the simplex partition sum $Z_i$ for all possible secondary structures of that sequence, and the corresponding duplex partition sums $Z_{ij}$ that result from folding a duplex involving two molecules $S_i$ and $S_j$. Note that duplex formation is concentration

dependent, and we therefore need to calculate the partition sum $\mathcal{Z}$ of the ensemble of sequences [4–6]. If each sequence is initially present in $n_i^0$ copies, and the ensemble after hybridization will contain $n_i$ simplex structures and $n_{ij}$ duplex structures, this partition sum can be written as:

$$\mathcal{Z} = \prod_i \frac{n_i^0!}{n_i! \prod_{j \leq i} n_{ij}!} Z_i^{n_i} \prod_{j \leq i} Z_{ij}^{n_{ij}}, \tag{1}$$

under the mass conservation constraint that each sequence be part of at most one complex at the same time:

$$n_i + 2n_{ii} + \sum_{j \neq i} n_{ij} = n_i^0. \tag{2}$$

The chemical equilibrium is obtained by minimizing the ensemble free energy $\mathcal{F} = -k_\mathrm{B} T \ln \mathcal{Z}$ with respect to the variables $n_i$ and $n_{ij}$, under the constraint Eq. (2). Even though in our relatively small system these variables are all small numbers, we can efficiently perform this calculation only in the thermodynamic limit, assuming rapid chemical equilibration due to the very fast hybridization kinetics [7] and the convective flow cycles encountered in the thermal trap. Hence, we switch to concentration variables $c_i = n_i/V$ in a volume $V$ (correspondingly for $c_i^0$ and $c_{ij}$).

Following Ref. [6], we now introduce Lagrange multipliers $\lambda_i$ (which are chemical potentials measured in units of $k_\mathrm{B}T$), and minimize $\mathcal{L} = \mathcal{F}/k_\mathrm{B}T + \sum_i \lambda_i (c_i^0 - c_i - 2c_{ii} - \sum_{j \neq i} c_{ij})$ instead. Using Stirling's formula, this requires finding the minimum of

$$\mathcal{L}(c, \lambda) = \sum_i \left[ c_i^0 (1 - \ln c_i^0 + \lambda_i) - c_i(1 - \ln c_i + \ln Z_i + \lambda_i) \right. \\ \left. - \sum_{j \leq i} c_{ij}(1 - \ln c_{ij} + \ln Z_{ij} + \lambda_i + \lambda_j) \right]. \tag{3}$$

The minimum is given by

$$c_i^* = Z_i e^{\lambda_i^*}, \quad c_{ij}^* = Z_{ij} e^{\lambda_i^* + \lambda_j^*}, \tag{4}$$

where the stationary values $\lambda^*$ for the chemical potentials are obtained from minimizing

$$f(\lambda) = -\mathcal{L}(c^*, \lambda) = \sum_i \left[ c_i^0 (\ln c_i^0 - 1 - \lambda_i) + Z_i e^{\lambda_i} + \frac{1}{2} \sum_j (1 + \delta_{ij}) Z_{ij} e^{\lambda_i + \lambda_j} \right]. \tag{5}$$

Even though this lower-dimensional problem is in principle not ill-conditioned [6], the minimization becomes numerically unstable for large systems on the order of 100 molecules with possibly very different hybridization energies. A stable code was obtained by using the

5

L-BFGS library [8] implementing the limited-memory Broyden-Fletcher-Goldfarb-Shanno algorithm [9], to obtain equilibrium values of $c^*$ that obey the mass conservation Eq. (2) within a relativ error of at most $10^{-4}$.

The probability $p_{L_i,K} = \frac{1}{2}[p_{i,K}+p_{i,K+1}]$ that enters the cleavage rate of bond $K$ of sequence $S_i$ involves the probabilities $p_{i,K,j,K'}$ that nucleotide $K$ is paired with another nucleotide at position $K'$ of sequence $S_j$, and therefore the probability $c_{ij}/c_i^0$ that sequence $S_i$ is actually part of the corresponding duplex:

$$p_{i,K} = \sum_{K'} p_{i,K,i,K'} \frac{c_i}{c_i^0} + \sum_{j,K'} p_{i,K,j,K'} \frac{c_{ij}}{c_i^0}. \tag{6}$$

The partition sum $Z_i$ of a sequence, and the duplex partition sums $Z_{ij}$ and corresponding base pairing probabilities $p_{i,K,j,K'}$ with all other sequences, are computed only once during the simulation, namely in the instant a sequence appears for the first time. For the computation of the effective cleavage rate $\beta_{L_i,K} = \beta_0(1 - p_{L_i,K})$, only the equilibrium concentrations $c_i$ and $c_{ij}$ need to be adjusted every time the sequence ensemble is modified. For this step, we only consider events involving sequences large enough to actually fold, i.e., we neglect the influence of changes in mono- and dinucleotide concentration.

Note that our scenarios operate at vastly different temperatures, which gives reason to question the quantitative accuracy of the RNA folding results. While the primary temperature dependence in the Boltzmann factors is correctly accounted for, the indirect dependence of the energy parameters used in the algorithm is captured only via a linear approximation around $T = 37\,°\mathrm{C}$. However, experimental RNA melting curves have been reproduced reasonably well over a wide range of temperatures [10], and we believe that small quantitative errors should not severely affect our results.

## III. DERIVATION OF THE STEADY-STATE LENGTH DISTRIBUTION

In the absence of sequence-specific cleavage rates, the sequence length distribution is identical to the cluster size distribution obtained in a simple aggregation-fragmentation process with a mass-independent aggregation rate $\alpha$ and a fragmentation rate $\beta L$ that is proportional to cluster size $L$, with random binary breakage. As a variation on the standard problems discussed in the literature, we also include monomer influx with rate $J$ and a length-dependent outflux $d_L$. For our parameter regimes, we expect that the aggregation-

fragmentation dynamics results in a nonequilibrium steady state length distribution $N_L$ (here we omit the angle brackets). It is obtained as the stationary solution of the following mass-balance equation:

$$\dot{N}_L = \alpha \sum_{K=1}^{L-1} N_K N_{L-K} - 2\alpha N_L \sum_{K=1}^{\infty} N_K - \beta(L-1)N_L + 2\beta \sum_{K=L+1}^{\infty} N_K + J_0 \delta_{L,1} - d_0 N_L e^{-\sqrt{L/L_c}}. \tag{7}$$

The first term on the right hand side describes the creation of a sequence of length $L$ from two fragments of sizes $K$ and $L-K$, while the next term models the ligation of sequence $S_L$ to any other sequence (the factor of 2 accounts for the correct counting of same-mass clusters). The third term corresponds to the breakage of sequence $S_L$ at any of its $L-1$ bonds, and the next term the production of a sequence of length $L$ as one of the two cleavage fragments of a longer sequence. The fifth and sixth terms, respectively, are monomer influx and the length-dependent outflux, where the square-root dependence in the exponential stems from the specific thermodiffusive behavior of polynucleotides in a thermal trap, and the crossover scale $L_c$ combines parameters such as the Soret coefficient, the trap geometry and the temperature difference across the trap. Because this complicated length dependence inhibits further analysis, we will set $L_c \to \infty$ in the following.

To proceed with the analysis, we perform the continuum limit $N_L \to N(L)$ in the rate equations Eq. (7):

$$\dot{N}(L) = \alpha \int_0^L dK\, N(K)N(L-K) - 2\alpha N(L) \int_0^\infty dK\, N(K) - \beta L N(L) \\ + 2\beta \int_L^\infty dK\, N(K) + J\delta(L) - d_0 N(L). \tag{8}$$

Now we introduce the moments

$$M_n = \int_0^\infty dL\, L^n N_L, \tag{9}$$

where $M_0 = \langle N_{\text{tot}} \rangle$ is the number of molecules, $M_1 = \langle L \rangle \langle N_{\text{tot}} \rangle$ is the total mass, and so forth. The rate equations for the moments are given by

$$\dot{M}_n = \alpha \sum_{k=0}^n \binom{n}{k} M_k M_{n-k} - 2\alpha M_0 M_n + \beta \left(\frac{2}{n+1} - 1\right) M_{n+1} + J_0 - d_0 M_n. \tag{10}$$

Even though the hierarchy of rate equations for the moments is not closed due to the

7

fragmentation term, the equations for the first two moments decouple from the rest:

$$\dot{M}_0 = -\alpha M_0^2 + \beta M_1 + J - d_0 M_0 \tag{11}$$

$$\dot{M}_1 = J - d_0 M_1. \tag{12}$$

Hence, we can easily obtain stationary solutions for the total number of molecules $\langle N_{\text{tot}} \rangle$ and their mean length $\langle L \rangle$:

$$\langle N_{\text{tot}} \rangle = \sqrt{\frac{d_0^3 + 4\alpha J(\beta + d_0)}{4\alpha^2 d_0}} - \frac{d_0}{2\alpha} \approx \sqrt{\frac{J(\beta + d_0)}{\alpha d_0}} \quad \text{if } J \gg d_0^3/(\alpha(\beta + d_0)), \tag{13}$$

$$\langle L \rangle = \frac{J}{d_0 \langle N_{\text{tot}} \rangle} \approx \sqrt{\frac{J\alpha}{d_0(\beta + d_0)}}. \tag{14}$$

Numerical studies indicate that the resulting length distribution is very similar to a Γ-distribution, which can be used in a moment closure approximation to compute higher moments [11]. For our parameter regime, the distribution is in fact close to exponential ($\langle \Delta L^2 \rangle \approx \langle L \rangle^2$).

We find that the thermal trap, through an outflux rate that drops with the exponential of the square root of sequence length, serves mainly to shift the distribution towards longer sequences. It does not, however, significantly affect the shape of tail of the distribution, because the dynamics of the longer molecules is mostly determined by their cleavage rate, which scales linearly with sequence length and thus quickly beats the outflux. In our simulations we kept $L_c = 10$ finite, because thermophoretic accumulation is essential to obtain nucleotides at reasonably high concentration in an experimental system. The above analysis suggests that the precise value of $L_c$ does not significantly affect our results.

## IV. DERIVATION OF THE AUTO- AND CROSS-CORRELATION FUNCTION

The master equation for the production and destruction of motifs of length $\ell$ and their complements is given by:

$$\partial_t p_{n,n^*} = k_+[p_{n-1,n^*} + p_{n,n^*-1}] + k_-[(n+1)h(\tfrac{n^*}{n+1})p_{n+1,n^*} + (n^*+1)h(\tfrac{n}{n^*+1})p_{n,n^*+1}] \\ - [2k_+ + k_-(nh(\tfrac{n^*}{n}) + n^*h(\tfrac{n}{n^*}))]p_{n,n^*}, \tag{15}$$

where $n$ and $n^*$ are the copy number of a motif and its complement, respectively, $k_+$ and $k_-$ are its birth and death rates, and $h(x)$ is the "hybridization function", which describes

the decrease in the death rate of a motif in terms of the probability of hybridization, which in turn is proportional to the number $x = n^*/n$ of complements per motif that are available for base pairing.

The dynamics of the mean $\langle n(t) \rangle = \sum_{n,n^*} n p_{n,n^*}(t)$ follows as

$$\partial_t \langle n \rangle = k_+ - k_- \langle n h(n^*/n) \rangle. \tag{16}$$

Assuming that $h(x)$ decreases only slowly from unity due to a small hybridization probability, we write

$$h(\tfrac{n^*}{n}) \approx 1 - |h'(0)| \tfrac{n^*}{n}, \tag{17}$$

which gives

$$\partial_t \langle n \rangle \approx k_+ - k_- \langle n \rangle + r \langle n^* \rangle, \tag{18}$$

where $r = k_- |h'(0)|$ is the apparent replication rate. Note that Eq. (15) is symmetric with respect to $n$ and $n^*$, and we can therefore directly infer the corresponding equation for $\langle n^* \rangle$. Conditional on the initial conditions $\langle n(0) \rangle = n_0$ and $\langle n^*(0) \rangle = n_0^*$, the solution of these two equations reads:

$$\langle n(t) \rangle_{n_0, n_0^*} = \frac{k_+}{k_- - r}\left(1 - e^{-(k_- - r)t}\right) + \frac{1}{2}(n_0 - n_0^*)e^{-(k_- + r)t} + \frac{1}{2}(n_0 + n_0^*)e^{-(k_- - r)t}. \tag{19}$$

The correlation functions $C_a(t)$ and $C_c(t)$ are defined as

$$C_a(t) = \langle n(t) n(0) \rangle = \sum_{n_0, n_0^*} n_0 \langle n(t) \rangle_{n_0, n_0^*} p^0_{n_0, n_0^*}, \tag{20a}$$

$$C_c(t) = \langle n(t) n^*(0) \rangle = \sum_{n_0, n_0^*} n_0^* \langle n(t) \rangle_{n_0, n_0^*} p^0_{n_0, n_0^*}, \tag{20b}$$

where $p^0_{n,n^*}$ is the steady-state solution of Eq. (15). All we actually need are the three steady-state averages $\langle n_0 \rangle = \langle n_0^* \rangle$, $\langle n_0^2 \rangle = \langle n_0^{*2} \rangle$ and $\langle n_0 n_0^* \rangle$, which are obtained from Eq. (15) by expanding the hybridization function as in Eq. (17):

$$\langle n_0 \rangle = \langle n_0^* \rangle = \frac{k_+}{k_- - r} \tag{21}$$

$$\langle n_0^2 \rangle = \langle n^{*2} \rangle = \frac{k_+}{k_- - r} \frac{k_-(k_- + k_+) + (k_+ - k_-)r}{k_-^2 - r^2} \tag{22}$$

$$\langle n_0 n_0^* \rangle = \frac{k_+}{k_- - r} \frac{k_-(k_+ + r) + (k_+ - r)r}{k_-^2 - r^2}. \tag{23}$$

Evaluating Eq. (20) gives Eq. (6) in the main text.

For a scenario with actual replication according to the reaction $n \xrightarrow{rn^*} n+1$, the master equation reads:

$$\partial_t p_{n,n^*} = k_+[p_{n-1,n^*} + p_{n,n^*-1}] + k_-[(n+1)p_{n+1,n^*} + (n^*+1)p_{n,n^*+1}] \\ + r[n^* p_{n-1,n^*} + n p_{n,n^*-1}] - [2k_+ + (k_- + r)(n+n^*)]p_{n,n^*}. \tag{24}$$

It is easy to check that this equation gives rise to the same expression Eq. (19) for $\langle n(t) \rangle$ as Eq. (15). However, the stationary second moments $\langle n_0^2 \rangle$ and $\langle n_0 n_0^* \rangle$ are slightly different:

$$\langle n_0^2 \rangle = \langle n_0^{*2} \rangle = \frac{k_+}{k_- - r} \frac{k_-(k_+ + k_-) + k_+ r}{k_-^2 - r^2} \tag{25}$$

$$\langle n_0 n_0^* \rangle = \frac{k_+}{k_- - r} \frac{k_-(k_+ + r) + k_+ r}{k_-^2 - r^2}. \tag{26}$$

The resulting correlation functions read:

$$C_{\mathrm{a/c}}(t) = \frac{k_+^2}{(k_- - r)^2} + \frac{k_- k_+ \mathrm{e}^{-(k_- - r)t}}{2(k_- - r)^2} \pm \frac{k_- k_+ \mathrm{e}^{-(k_- + r)t}}{2(k_-^2 - r^2)}. \tag{27}$$

The time dependence, given through Eq. (19), is clearly the same as that of the correlation functions of Eq. (15), and the amplitudes are identical to those of Eq. (6) in the main text to first nonzero order in $r$:

$$C_{\mathrm{a}}(0) - C_{\mathrm{a}}(\infty) = \frac{k_+}{k_-} + \mathcal{O}(r), \tag{28}$$

$$C_{\mathrm{c}}(0) - C_{\mathrm{c}}(\infty) = \frac{k_+ r}{k_-^2} + \mathcal{O}(r^2). \tag{29}$$

---


\end{document}